\documentclass[prb, preprint, twocolumn, amsmath, amssymb, reprint, superscriptaddress]{revtex4-1}

\usepackage{graphicx,verbatim}
\usepackage{dcolumn}
\usepackage{bm}
\usepackage{sidecap}
\usepackage{braket}
\usepackage{color}
\usepackage{subcaption}
\usepackage{tikz}
\newcommand*\circled[1]{\tikz[baseline=(char.base)]{
            \node[shape=circle,draw,inner sep=2pt] (char) {#1};}}

\begin{document}

\title{ How an electrical current can stabilize a molecular nanojunction }

\author{A. Erpenbeck}
\affiliation{
	Department of Physics, University of Michigan, Ann Arbor, Michigan 48109, USA
	}
	
\author{Y.\ Ke}
\affiliation{
    Institute of Physics, University of Freiburg, \\
    Hermann-Herder-Strasse 3, 79104 Freiburg, Germany
    }

\author{U.\ Peskin}
\affiliation{
    Schulich Faculty of Chemistry, Technion-Israel Institute of
    Technology, Haifa 32000, Israel
    }

\author{M.\ Thoss}
\affiliation{
    Institute of Physics, University of Freiburg, \\
    Hermann-Herder-Strasse 3, 79104 Freiburg, Germany
    }

\date{\today}

\begin{abstract}
		The stability of molecular junctions under transport is of the utmost importance for the field of molecular electronics.
		This question is often addressed within the paradigm of current-induced heating of nuclear degrees of freedom or current-induced forces acting upon the nuclei.
		At the same time, an essential characteristic of the failure of a molecular electronic device is its changing conductance -- typically from a finite value for the intact device to zero for a device that lost its functionality.
		In this publication, we focus on the current-induced changes in the molecular conductance, which are inherent to molecular junctions at the limit of mechanical stability.
		We employ a numerically exact framework based on the hierarchical equations of motion approach, which treats both electronic and nuclear degrees of freedom on an equal footing and does not impose additional assumptions.  
		Studying generic model systems for molecular junctions with dissociative potentials for a wide range of parameters spanning the adiabatic and the nonadiabatic regime, 
		we find that molecular junctions that exhibit a decrease in conductance upon dissociation are more stable than junctions that are more conducting in their dissociated state.
		This represents a new mechanism that stabilizes molecular junctions under current.
		Moreover, we identify characteristic signatures in the current of breaking junctions related to the interplay between changes in the conductance  and the nuclear configuration and show how these are related to properties of the leads rather than characteristics of the molecule itself.
\end{abstract}
\maketitle

\section{Introduction}

    Assessing the stability of molecular junctions under the influence of a current is of fundamental importance in the field of molecular electronics. It ultimately determines their prospect as next generation electronic devices. Moreover, investigations of the stability of molecular junctions and the underlying processes leading to their failure reveal information on the molecule itself and on the underlying nonequilibrium physics.
    
    The most prominent notion for considering molecular stability is the concept of current induced heating of vibrational modes.
    Several experimental and theoretical studies confirmed the existence of current-induced vibrational excitations,\cite{Ioffe2008, Schulze2008, Schulze2008_2, deLeon2008, Huettel2009, Rainer2009, Rainer2010, Ward2010, Rainer2011, Rainer2011b, Franke2012, Schinabeck2016, Schinabeck_Hierarchical_2018, Bi_Electron_2020} their influence on the conductance properties,\cite{Galperin_Vib_Effects, Galperin_Inelastic_2004, Pop_Negative_2005, Galperin_Resonant_2006, Koch2006, Koch_Theory_2006, Donarini_Dynamical_2006, Leijnse2008, Park_Self_2011, Volkovich2011, Rainer2012, Erpenbeck2015, Erpenbeck2016, Popp_Thermoelectricity_2021, Kaspar_Nonadiabatic_2022}
    and their impact on the mechanical instability of the molecular junction.\cite{Venkataraman2015, Sabater2015, Venkataraman2016, Capozzi2016} This is in line with the fact that molecular junctions are seldom stable beyond a bias voltage of $\sim 1-2$ V.\cite{Schulze2008, Sabater2015} 
    In our recent publications Refs.\ \onlinecite{Erpenbeck_Current_2020, Ke_Unraveling_2021}, 
    we studied different dissociation mechanisms in great detail and assessed the importance of current-induced vibrational excitation.
    Beyond the paradigm of current-induced heating, a variety of studies approached the question of the stability of molecular junctions based on a classical description of the nuclear degrees of freedom (DOFs) and current-induced forces acting on them.\cite{Dzhioev2011, Dzhioev2013, Pozner2014, Erpenbeck_dissociation_2018, preston2021firstpassage}
    Moreover, recent findings suggest that external driving and increased temperature can stabilize molecular junctions.\cite{Hartle_Cooling_2018, Kuperman_Mechanical_2020, Preston_Cooling_2020}

    \begin{figure}[tb]
        \vspace*{0.3cm}
        \raggedright a)
        \begin{minipage}[c]{0.5\textwidth}
	    \vspace*{-0.2cm}
	    \hspace*{-0.3cm}
            \includegraphics[width=0.9\textwidth]{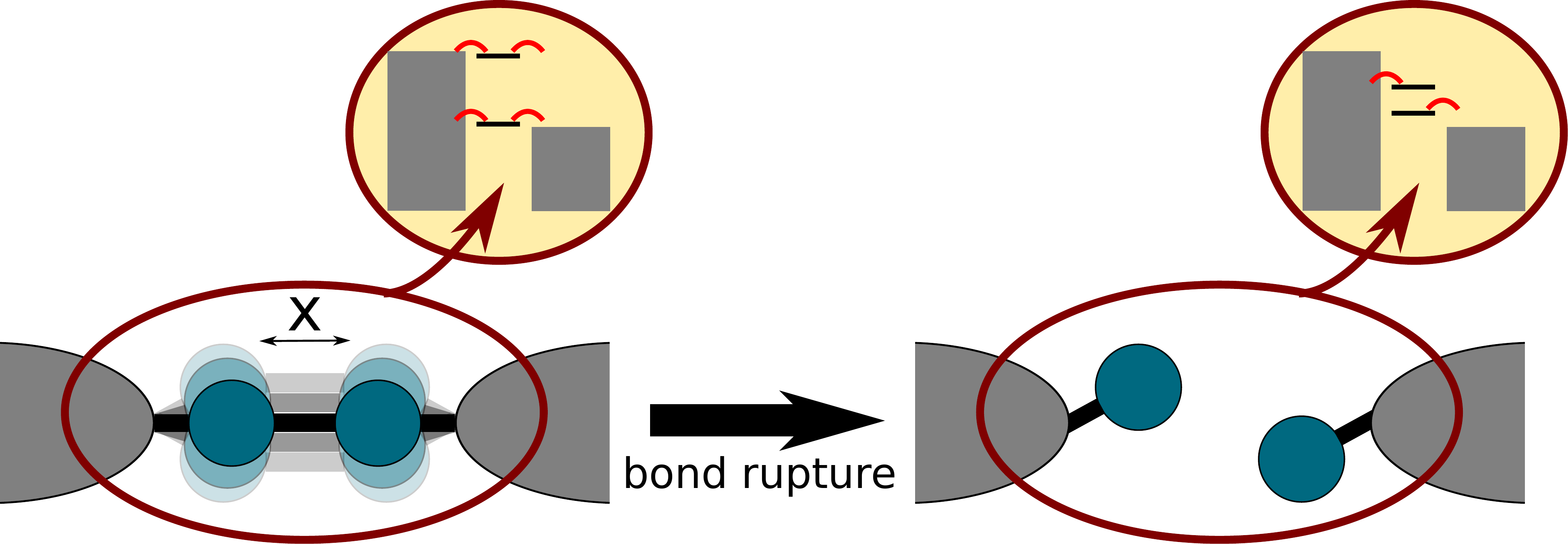}
        \end{minipage}\\
        \vspace*{1cm}
        \raggedright b)
        \begin{minipage}[c]{0.5\textwidth}
	    \vspace*{-0.7cm}
	    \hspace*{-0.3cm}
            \includegraphics[width=0.9\textwidth]{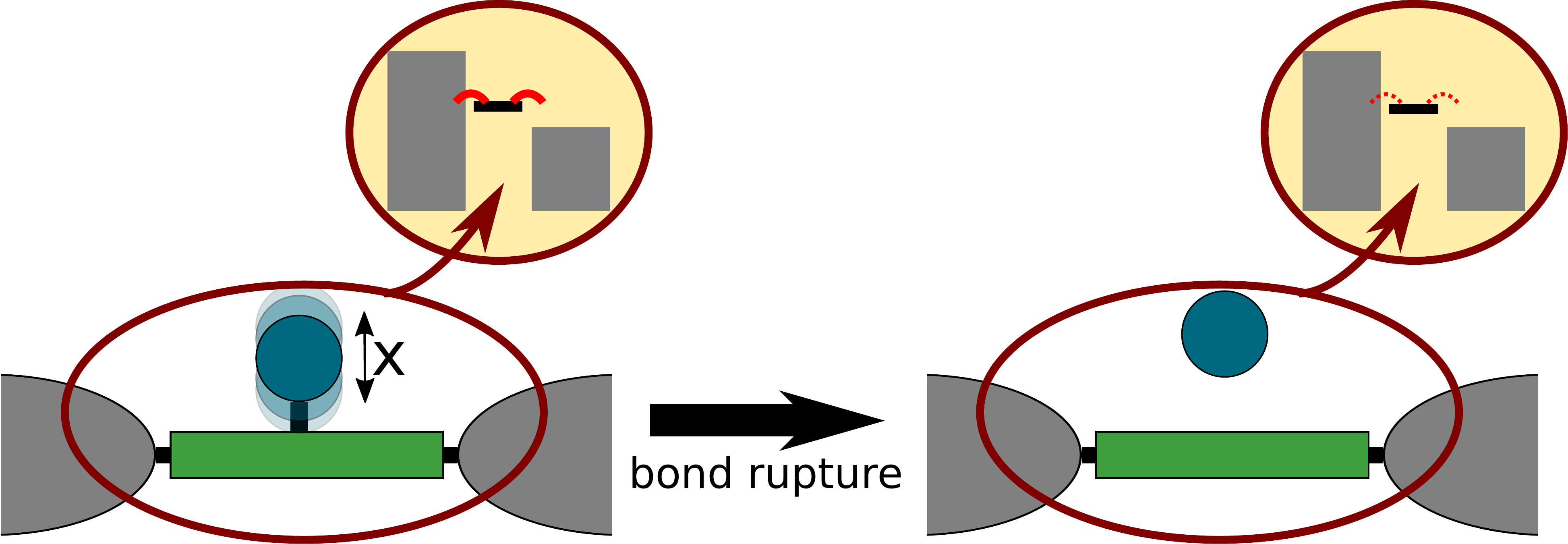}
        \end{minipage}
        \captionsetup{font=small,labelfont=bf, justification=centerlast, format=plain}
        \caption{ \bf \scriptsize
            Representative setups of non-rigid molecular junctions. The blurry parts highlight the current-induced structural movement of the molecular junction. The associated representation in terms of molecular eigenstates is provided in the yellow circles, whereby red arcs represent the coupling between molecular states (black line) and the leads (gray shaded areas).
            a: Flexible atomic chain, which is representative for destructive current-induced bond-rupture.
            b: Molecular junction where the conductance is influenced by a side-group.\cite{Erpenbeck_dissociation_2018, Erpenbeck_2019_HQME, Erpenbeck_Current_2020}
            }	
        \label{fig:setups}
    \end{figure}
    
    All the studies mentioned above focus on intrinsic properties of the molecule itself to assess its stability in a junction under transport. 
    A fundamental aspect of molecular junctions, especially when close to current-induced mechanical instability, is the change in their conductance behavior. If a molecular junction breaks, its conductance changes from a finite value to zero as the molecule starts to fall apart.
    Another possibility is that current-induced bond-ruptures within the molecule itself change the internal structure of the molecule, leading to a massive change in its conductance, thus counteracting further current-induced modulations within the molecule in the junction. 
    This change in conductance for a molecular junction undergoing structural changes was also observed experimentally.\cite{Arielly_Picosecond_2017}
    
    Working in the eigenbasis of the molecule, a change in the conductance is reflected in the coupling between the molecular states and the leads, which needs to depend on the nuclear DOFs to allow for current-induced changes in the conductance.
    To make this more apparent, consider a tight-binding chain consisting of two sites as depicted in Fig.\ \ref{fig:setups}a. 
    In this case, the distance between the two sites represents the generalized nuclear coordinate which can be influenced by a current and which mediates the tunneling rate between the two sites. 
    Assuming that a large current breaks the junction, i.e.\ the distance between the two sites diverges (right hand side of Fig.\ \ref{fig:setups}a), it is possible to describe this scenario within the molecular eigenstates using nuclear coordinate dependent energies and molecule-lead coupling strength (yellow shaded areas in Fig\ \ref{fig:setups}a).
    Another example is a nondestructive system depicted in Fig.\ \ref{fig:setups}b which was studies in Refs.\ \onlinecite{Erpenbeck_dissociation_2018, Erpenbeck_2019_HQME, Erpenbeck_Current_2020, Ke_Unraveling_2021}.
    Here, the molecule consists of a backbone and a side-group. If the side-group detaches due to an imposed current (right hand side of Fig.\ \ref{fig:setups}b), this can, for example, result in the destruction of a $\pi$-conjugation within the molecular backbone leading to a decreasing conductance. Again, this scenario can be described by nuclear coordinate dependent energies and molecule-lead coupling strength.
    
    Despite the universality of a change in conductance upon current-induced changes in the nuclear configuration in molecular junctions, only a few selected publications have explicitly allowed for nuclear-coordinate dependent molecule-lead coupling strengths.\cite{Maytal2007, Dou_Universality_2017, Dou_Born_2017, Dou_Electronic_2017, Coffman_When_2018, Hopjan_Molecular_2018, Erpenbeck_dissociation_2018, Erpenbeck_2019_HQME, Erpenbeck_Current_2020, Preston_Current_2020, Ke_Unraveling_2021, Ke_Nonequilibrium_2022}
    In the present paper, we study the effect of a conductance depending on the nuclear DOFs, its back-action on the nuclear dynamics, and its implications for the stability of molecular junctions.
    
    The remainder of this paper is organized as follows: 
    In Sec.~\ref{sec:model}, we introduce our model system, review the methodology which we use to obtain a numerically exact description for the system, and introduce our main observables.
    Our results are presented in Sec.~\ref{sec:results}, whereby we find distinctively different behavior in the weak-coupling regime (Sec.~\ref{sec:weak-coupling}) and the strong coupling regime where a classical interpretation of the nuclear DOFs becomes meaningful (Sec.~\ref{sec:strong-coupling}).
    We summarize and conclude in Sec.~\ref{sec:conclusion}.

\section{Model and method}\label{sec:model}

	Molecular junctions comprise electronic and nuclear DOFs. The leads act as electronic reservoirs and enable transport. The Hamiltonian for the molecular junction is
	$H = H_{\text{S}} + H_{\text{B}} + H_{\text{SB}}$, 
	where $H_{\text{S}}$ is the Hamiltonian of the system, $H_{\text{B}}$ the Hamiltonian describing the electronic baths, and the coupling between them is given by $H_{\text{SB}}$. 
	Quite generally, the individual parts are assumed to be of the form
	\begin{subequations}
	\begin{eqnarray}
		H_{\text{S}}	&=& 	\sum_\alpha \frac{p_\alpha^2}{2m_\alpha} + \sum_{\mu\nu} \epsilon_{\mu\nu}(\textbf{x}) \ d_\mu^\dagger d_\nu ,  \\
		H_{\text{B}}	&=&	\sum_{k\in \text{L/R}} \epsilon_k c_k^\dagger c_k, \label{eq:H_B}\\
		H_{\text{SB}}	&=&	\sum_{k\in\text{L/R}}\sum_{\mu} \left( V_{\mu k}(\textbf{x}) c_k^\dagger d_\mu  + \text{h.c.} \right) , \label{eq:H_coupl}
	\end{eqnarray}
	\end{subequations}
	where $\textbf{x}$ and $\textbf{p}$ are the position and momentum of the nuclear DOFs with associated mass $m_\alpha$. $d_\mu^\dagger$ and $d_\mu$ are the electronic creation and annihilation operators associated to the electronic state $\mu$ of the molecule.
	The leads are modeled as noninteracting baths, where $c_k^{\dagger}/c_k$ denote the electronic creation/annihilation operator associated to state $k$ with energy $\epsilon_k$ in the left or right lead (L/R).
	The coupling between the molecule and the leads in Eq.~(\ref{eq:H_coupl}) enables transport across the molecule and gives rise to the spectral density 
	\begin{eqnarray}
		\Gamma_{\text{L/R} \nu\nu'}(\epsilon, \textbf{x}) = 2\pi\sum_{k\in \text{L/R}} V_{\nu k}(\textbf{x}) V_{\nu'k}^*(\textbf{x}) \delta(\epsilon-\epsilon_k) \ .  \label{eq:spectral_func}
	\end{eqnarray}
	As mentioned before, the dependence of the molecule-lead coupling $V_{\mu k}$ on the nuclear DOFs $\textbf{x}$ is essential to model situations where the molecular conductance is influenced by nuclear reorganization; an effect that is fundamental to current-induced device failure. Moreover, it introduces a back-action mechanism for the current, i.e.\ the current drives the nuclei away from their equilibrium position, which influences the conductance of the molecule and therefore the current that was driving the nuclei in the first place.

	For the numerical investigations presented below, we employ a numerically exact framework which treats electronic and nuclear DOFs on an equal footing and does not impose additional assumptions. This framework was used before to study current-induced dissociation in molecular junctions.\cite{Erpenbeck_2019_HQME, Erpenbeck_Current_2020, Ke_Unraveling_2021}
	In the following, we restrict ourselves to naming the basic ingredients for this methodology.
	The method is presented and discussed in detail in Ref.~\onlinecite{Erpenbeck_2019_HQME}. For technical aspects concerning the application to dissociative molecular systems, we refer to Refs.~\onlinecite{Erpenbeck_Current_2020, Ke_Unraveling_2021}.
	
	Generally, the framework employed in this work is based on the hierarchical equations of motion approach (HEOM; also abbreviated as HQME).\cite{Tanimura_2020_J.Chem.Phys._p20901, Tanimura1989, Tanimura2006, Jin_2007_J.Chem.Phys._p134113, Jin_2008_J.Chem.Phys._p234703, Zheng_2009_J.Chem.Phys._p164708, Yan_2014_J.Chem.Phys._p54105, Ye_2016_WIREsComputMolSci_p608, Haertle2013a, Haertle2015, Schinabeck2016, Erpenbeck_Extending_2018}
	This reduced density matrix scheme describes the dynamics of a quantum system influenced by an environment. It provides the exact result upon the solution of a large set of coupled differential equations, which are truncated in a systematic fashion until the desired precision is reached. For details and explicit expressions for the hierarchical equations, we refer to the recent review article Ref.~\onlinecite{Tanimura_2020_J.Chem.Phys._p20901}.
	For the scope of this work, we consider a molecule which is initially in its electronic and vibrational ground state and which has no initial correlations with the leads, and determine its dynamics upon propagating the differential equation of the HEOM method using a fourth order Runge-Kutta scheme.
	Some additional information on non-constant molecule-lead coupling strengths,\cite{Rahman_Non_2018} and how they can be incorporated in the HEOM scheme, are given in Ref.~\onlinecite{Erpenbeck_Hierarchical_2019}.
	
	The vibrational DOF is represented within the discrete variable representation.\cite{tannor2007introduction, Colbert_A_1992}
	In order to account for the semi-infinite nature of a dissociative problem, we introduce a complex absorbing potential (CAP) and an additional Lindblad-like source term. Thereby, the Lindblad term avoids problems with the conservation of particle number associated with the CAP, and allows for the definition of a dissociation probability. The details of which can be found in Ref.~\onlinecite{Erpenbeck_2019_HQME}.
	Beyond the dissociation probability, the electronic current and the force acting on the nuclear DOF are the observables of interest for the scope of the present work. Their calculation within the HEOM framework is detailed in Refs.~\onlinecite{Erpenbeck_dissociation_2018, Erpenbeck_Current_2020}.
	In particular, the electronic current between the molecule and lead $K$ is given by
	\begin{eqnarray}
		\hspace*{-0.5cm} 
		I_K 
		&=& \frac{ie}{\hbar^2} \hspace*{-0.2cm}\sum_{\mu; k\in K \atop p\in \text{poles}}\hspace*{-0.2cm} \text{Tr}_{\text{S}} \left(  V_{\mu k}(\mathbf{x})\left( \rho^{(1)}_{kp+}d_\mu - d_\mu^\dagger\rho^{(1)}_{kp-} \right) \right), \label{eq:current}
	\end{eqnarray}
	and we employ the notion of the force generated by the non-constant molecule-lead coupling,
	\begin{eqnarray}
		\hspace*{-0.5cm} 
		\mathbf{F}_{\text{SB}}
		&=& - m \text{Tr}_{\text{S+B}} \left( \varrho \frac{\partial H_{\text{SB}}}{\partial \mathbf{x}} \right) \nonumber \\ 
		&=& - m \hspace*{-0.2cm}\sum_{\mu; k\in \text{L/R} \atop p\in \text{poles}}\hspace*{-0.2cm} \text{Tr}_{\text{S}} \left(  \frac{\partial V_{\mu k}(\mathbf{x})}{\partial \mathbf{x}}\left( \rho^{(1)}_{kp+}d_\mu + d_\mu^\dagger\rho^{(1)}_{kp-} \right) \right). \label{eq:force}
	\end{eqnarray}
	Here, $\varrho$ is the density matrix of the molecule and the leads, $\text{Tr}_{\text{S+B}}$ denotes the trace over the DOFs of the molecule and the leads, $\text{Tr}_{\text{S}}$ is the trace over the molecular DOFs (electronic and vibrational), and $\rho^{(1)}_{kp\pm}$ are first-tier auxiliary density operators associated with lead state $k$ and pole $p$.
	We remark the similarity between the expressions for the force $\mathbf{F}_{\text{SB}}$ in Eq.\ (\ref{eq:force}) and for the electronic current in Eq.\ (\ref{eq:current}), both of which can be expressed solely in terms of first-tier auxiliary density operators. 
	Hence, $\mathbf{F}_{\text{SB}}$ is related to the electrons exchanged between the molecule and the leads, weighted by the derivative of the molecule-lead coupling strength with respect to the nuclear coordinates. 
	We mention that the trace over the first-tier density operators in Eq.\ (\ref{eq:force}), that is $\text{Tr}_{\text{S}} \left(\rho^{(1)}_{kp+}d_\mu + d_\mu^\dagger\rho^{(1)}_{kp-} \right)$, is found to be predominantly negative for the systems studied below, such that a molecule-lead coupling strength $V_{\mu k}(\mathbf{x})$ that decreases with increasing $\mathbf{x}$ will result in a force acting in the direction of smaller nuclear coordinate values.
	We note in passing that other authors have already analyzed the force stemming from a nuclear-coordinate dependent molecule-lead coupling strength, and that under certain conditions, issues were reported when employing the wide-band limit and a coupling to the environment depends on the nuclear DOFs.\cite{Dou_Molecular_2016}

\section{Results}\label{sec:results}
		\begin{figure}[tb!]
				\raggedright a)\\
				\raggedleft
				\vspace*{-0.3cm}
				\includegraphics[width=0.46\textwidth]{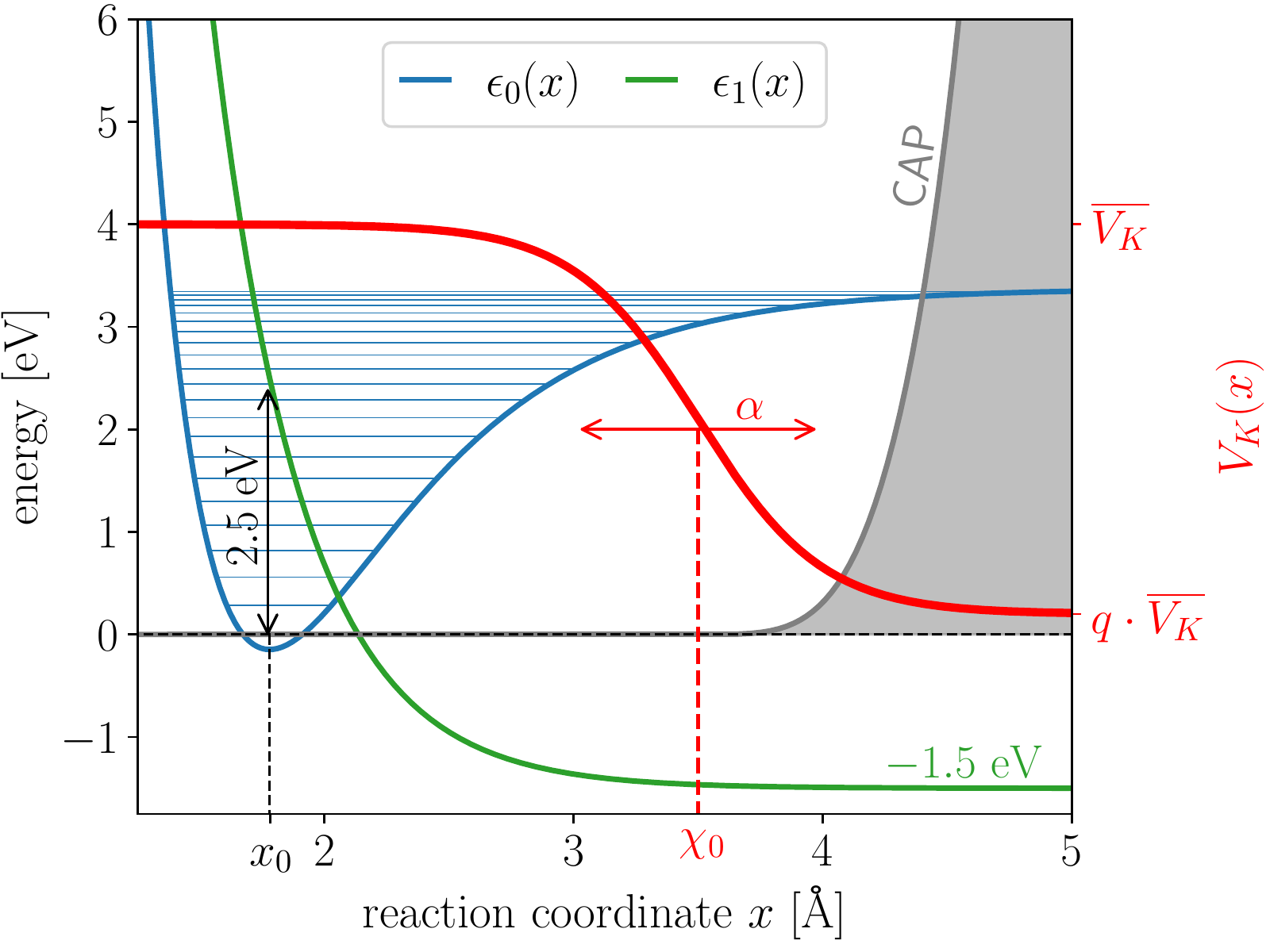}\\
				\raggedright b)\\
				\raggedleft
				\vspace*{-0.3cm}
				\includegraphics[width=0.46\textwidth]{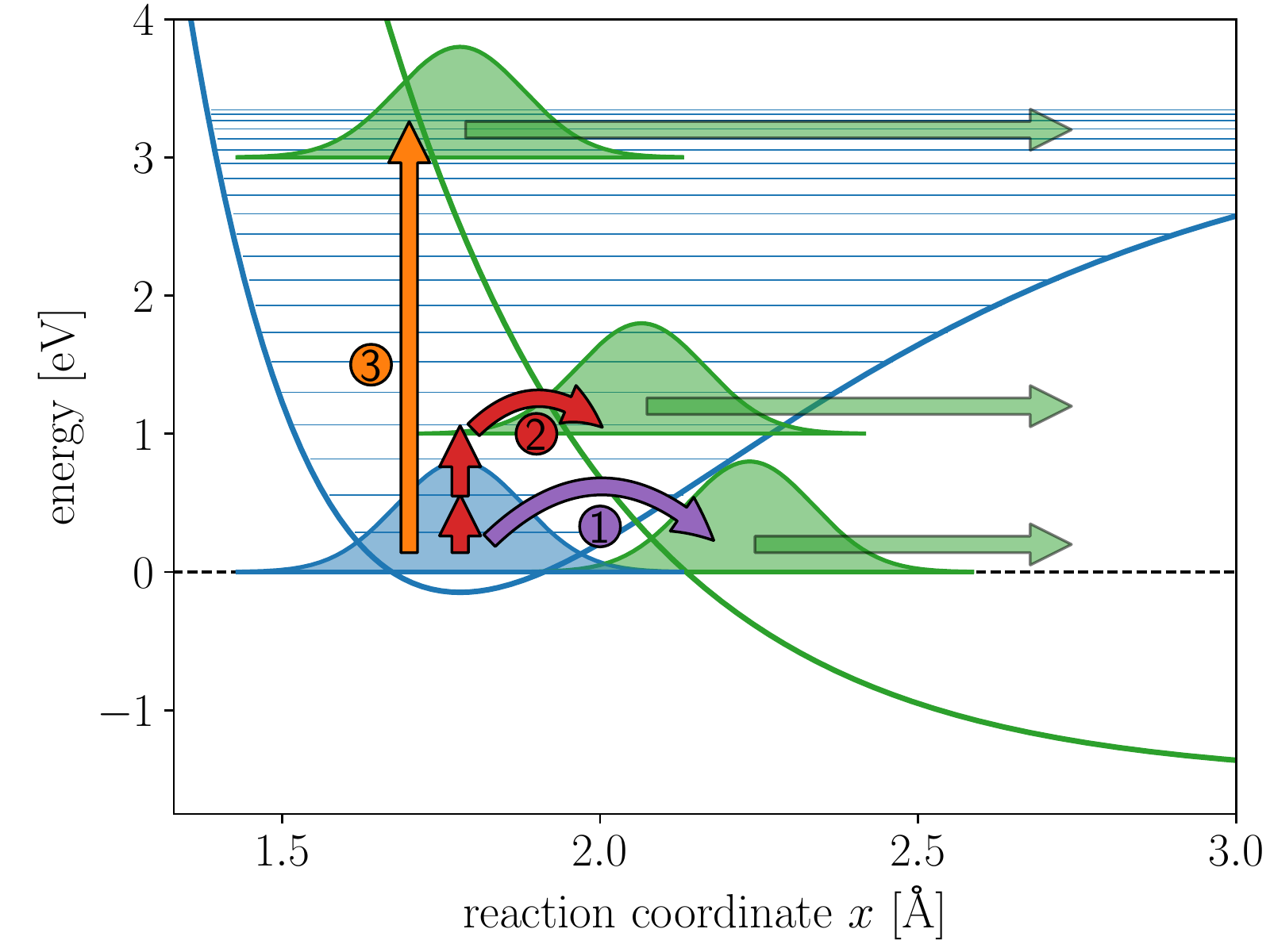}
			\captionsetup{font=small,labelfont=bf, justification=centerlast, format=plain}
			\caption{ \bf \scriptsize
					a: PESs used to model the bond between the backbone and the side-group of the molecular bridge in the neutral ($\epsilon_0(x)$) and the charged state ($\epsilon_1(x)$). The red line visualizes the dependence of the molecule-lead coupling $V_{K}(x)$ on the nuclear coordinate.
					b: Graphical representation of three different mechanisms leading to dissociation, which provide an explanation for the diverse timescales for the dissociation process. 
				}
			\label{fig:sketch}
		\end{figure}
		
	In the present work, we consider a minimal model of reduced complexity.
	We focus on non-destructive current-induced bond rupture in single-molecule junctions, which represents a general platform for studying the effect of current-induced conductance changes.
	We study the system depicted in Fig.~\ref{fig:setups}b, which consists of a side group which detaches from a backbone upon the influence of a current. 
	Our minimal model includes a single electronic state and a single one-dimensional reaction coordinate $x$ described by the Hamiltonian
	\begin{eqnarray}
        H_{\text{S}}	&=&	\frac{p^2}{2m} + \epsilon_0(x) dd^\dagger + \epsilon_1(x) d^\dagger d .
	\end{eqnarray}
	The molecule described by this model can be in a neutral state (electronic state is unpopulated) or in an anionic state (electronic state is populated).
	The potential energy surface (PES) of the neutral molecule is assumed to be a binding Morse-potential,
	while the PES of the charged molecule adopts a non-binding form,
	\begin{subequations}
        \begin{eqnarray}
            \epsilon_0(x) 	&=&	D_e \cdot\left( e^{-a(x-x_0)} -1 \right)^2 + c , \label{eq:V_0} \\ 
            \epsilon_1(x)   &=&	D_1 \cdot e^{-a' (x-x_0')}  + V_\infty. 
        \end{eqnarray}
	\end{subequations}
	In the numerical results presented below, we have used the parameters $m=1$ amu (atomic mass units), 
	$D_e = 3.52$ $\text{eV}$,  
	$x_0=x_0'=1.78$ \AA, 
	$a=1.7361$~\AA$^{-1}$, 
	$D_1=4.0$ $\text{eV}$, 
	$a'=2.758$ \AA$^{-1}$. 
	The parameter $c= -147$~meV is chosen such that the nuclear ground-state has the energy $0$ eV. $V_\infty$ is set to $-1.5$ eV. These potentials are visualized  in Fig. \ref{fig:sketch}a.
	Moreover, unless stated otherwise, the leads are modeled by a Lorentzian spectral density of width $W=100$~eV.
	In the calculations reported below, we assume that both leads have the same temperature $T=300$ K and that the bias voltage $\Phi$, defined as the difference between the chemical potentials $\Phi=\mu_{\text{L}}-\mu_{\text{R}}$, drops symmetrically such that $\mu_{\text{L}} = -\mu_{\text{R}}$.
	We note that this system was studied in in our previous publications Refs.\ \onlinecite{Erpenbeck_dissociation_2018, Erpenbeck_2019_HQME, Erpenbeck_Current_2020} in order to identify the basic mechanisms of current-induced bond rupture in molecular junctions; a sketch of the corresponding dissociation mechanisms is provided in Fig.~\ref{fig:sketch}b.
    In the weak molecule-lead coupling regime, dissociation is the result of the nuclear wave-function leaking into the classically forbidden regime. This mechanism is sketched as $\circled{1}$ in Fig.\ \ref{fig:sketch}b and can be enhanced by prior current-induced excitation of the nuclear DOF as depicted as process $\circled{2}$ in Fig.\ \ref{fig:sketch}b.
    The timescale for dissociation in this regime is long, as the mechanism for dissociation is essentially a tunneling phenomenon.
    In the strong coupling regime, the wave-packet remains compact while it propagates under the influence of the PESs. In this regime, the timescale for dissociation is relatively fast and an interpretation of the nuclear dynamics in terms of classical forces acting on the nuclear wave-packet provides the correct intuition.
	In the present manuscript, we extend these earlier findings by a detailed analysis of the influence of a nuclear-coordinate dependent molecule-lead coupling strength on the stability of a molecular junction.

	The central part of our investigation is the molecule-lead coupling strength, which depends on the nuclear coordinate, and which determines the conductance of the molecule. 
	Here, we assume that the molecule-lead coupling has the form
	\begin{eqnarray}
		V_{k}(x)    &=&    \overline V_{k} \cdot \left( \frac{1-q}{2} \left[ 1-\tanh\left(\frac{x-\chi_0}{\alpha} \right) \right] + q \right) , \ \label{eq:def_mol_lead_coupling_strength}
	\end{eqnarray}
	which allows for a smooth switching of the coupling from $\overline V_{k}$ for the intact junction, to $q\overline V_{k}$ for the dissociated junction. The switching process takes place in a range of $x$-values of extent $\alpha$ centered around $\chi_0$. This form of $V_{k}(x)$ is also depicted in Fig.~\ref{fig:sketch}a.
	As $\overline V_{k}$ sets the time-scale for the electronic dynamics, we use it to distinguish between the weak and strong coupling regime. 
	We use $q$ to differentiate between junctions whose conductivity is diminished ($q<1$) or increased ($q>1$) upon dissociation. The limiting case of a $x$-independent molecule-lead coupling is recovered for $q=1$.

	In the following, we discuss the dissociative behavior and the dynamics of the model introduced above.
	Thereby, we distinguish between two different regimes that show different physical behavior: 
	(i) the non-adiabatic regime, also known as weak molecule-lead coupling limit, where the electronic motion is slow compared to the nuclear timescales and 
	(ii) the adiabatic regime, or strong coupling regime, where the electronic dynamics is much faster than the nuclear motion. 
	Based on our previous works,\cite{Erpenbeck_Current_2020, Ke_Unraveling_2021} it is known that the dissociative mechanisms in these two regimes are distinctly different.
	Note that for all data presented below, we have tested for convergences with respect to all numerical parameters such that we are exclusively working with numerically exact results.

\subsection{Weak-coupling limit -- tunneling regime}\label{sec:weak-coupling}

	We start our analysis with the regime of small to intermediate bias voltages and small to moderate molecule-lead coupling strengths.
	In this regime, the dissociation is a result of the nuclear wave-function leaking into the classically forbidden regime, i.e.\ a tunneling phenomenon. The corresponding mechanism is sketched as $\circled{1}$ in Fig.\ \ref{fig:sketch}b.\cite{Erpenbeck_Current_2020} This process can be proceeded by current-induced vibrational heating ($\circled{2}$ in Fig.\ \ref{fig:sketch}b), which is of particular importance for systems displaying a pronounced non-resonant transport regime,\cite{Erpenbeck_Current_2020} or for systems with multiple attractive PESs.\cite{Ke_Unraveling_2021}
	
		\begin{figure}[htb!]
				\raggedright a)\\
				\raggedleft
				\vspace*{-0.6cm}
				\includegraphics[width=0.46\textwidth]{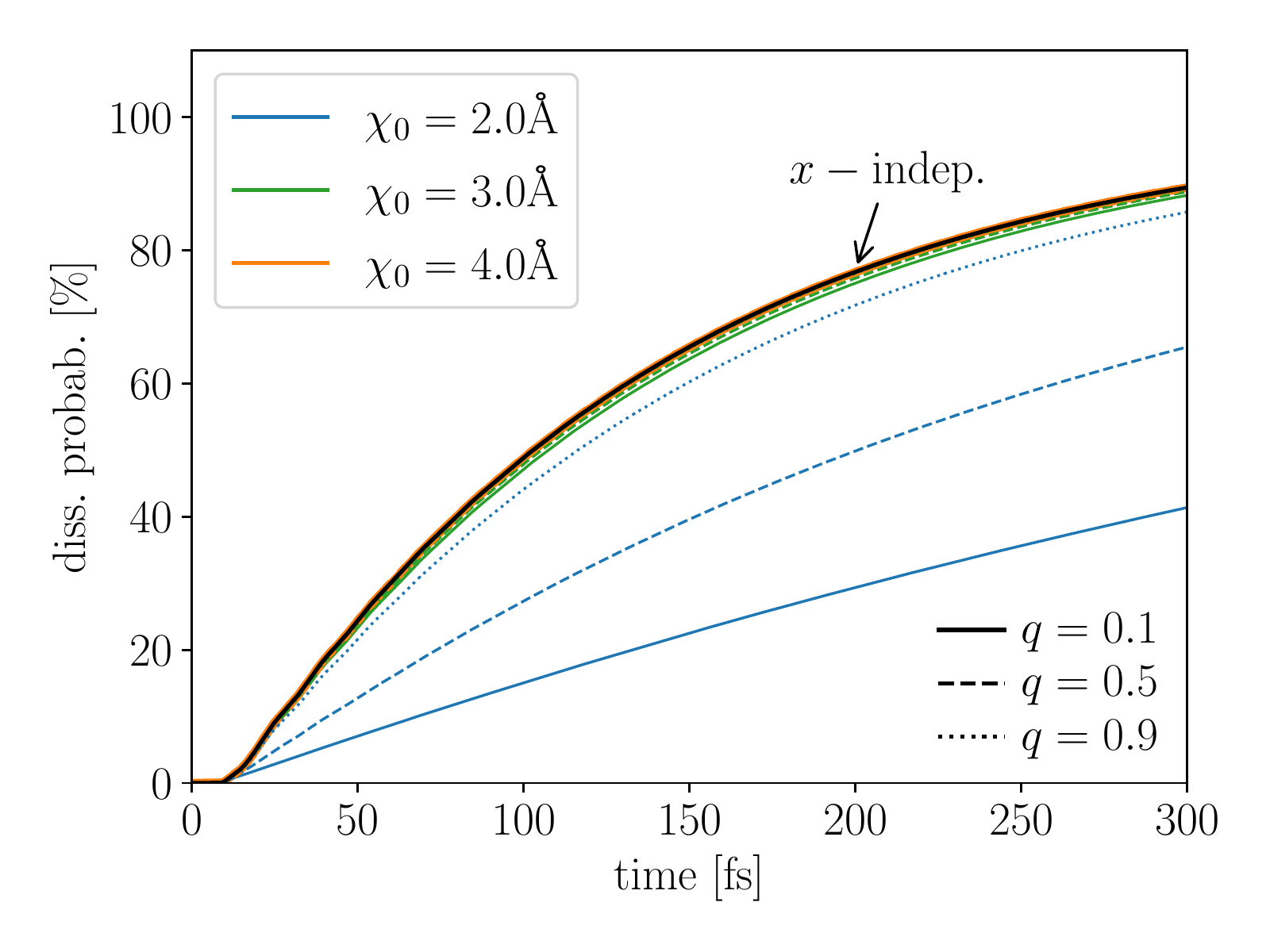}\\
				\raggedright b)\\
				\raggedleft
				\vspace*{-0.6cm}
				\includegraphics[width=0.46\textwidth]{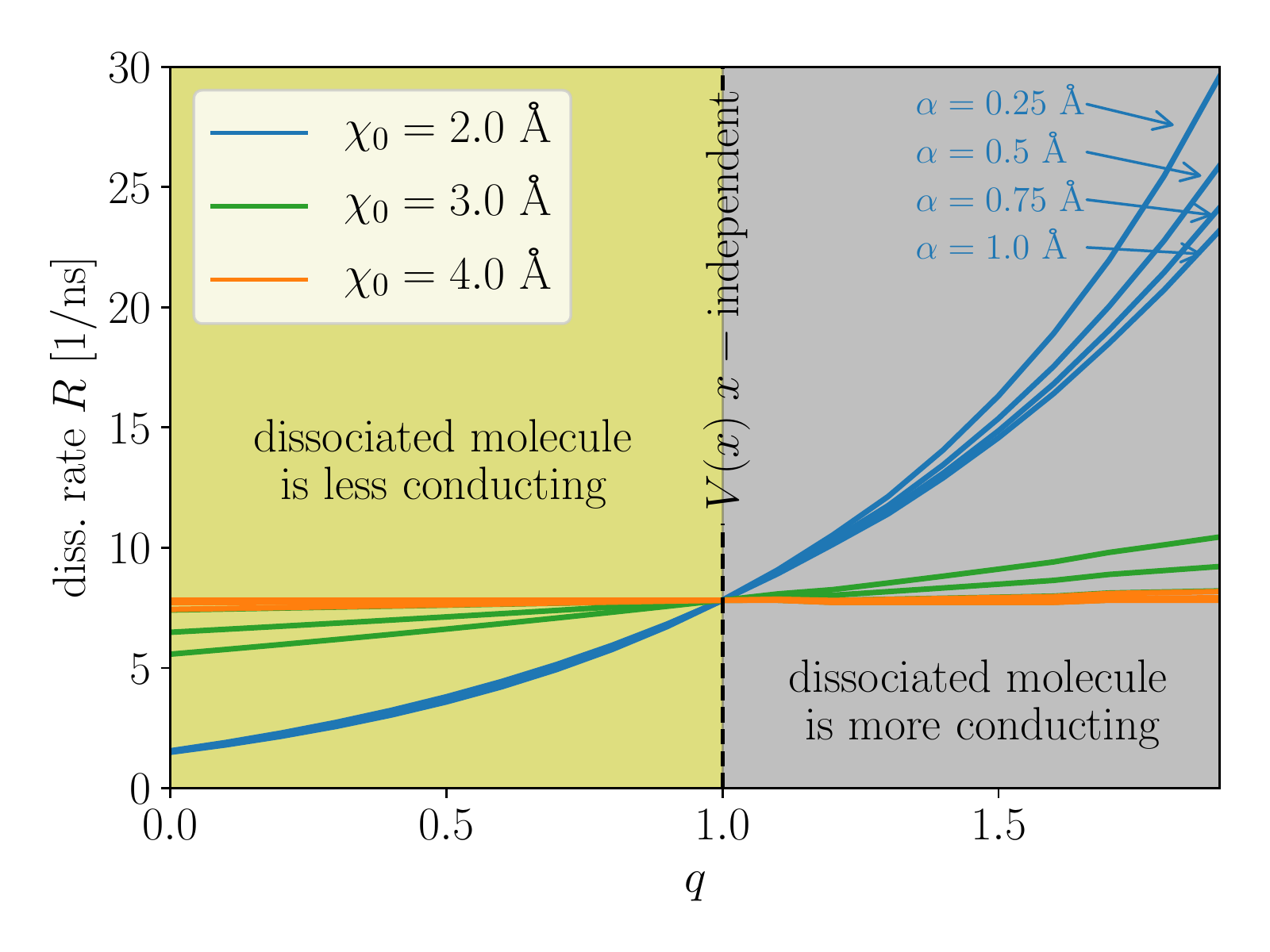}\\
				\raggedright c)\\
				\raggedleft
				\vspace*{-0.6cm}
				\includegraphics[width=0.46\textwidth]{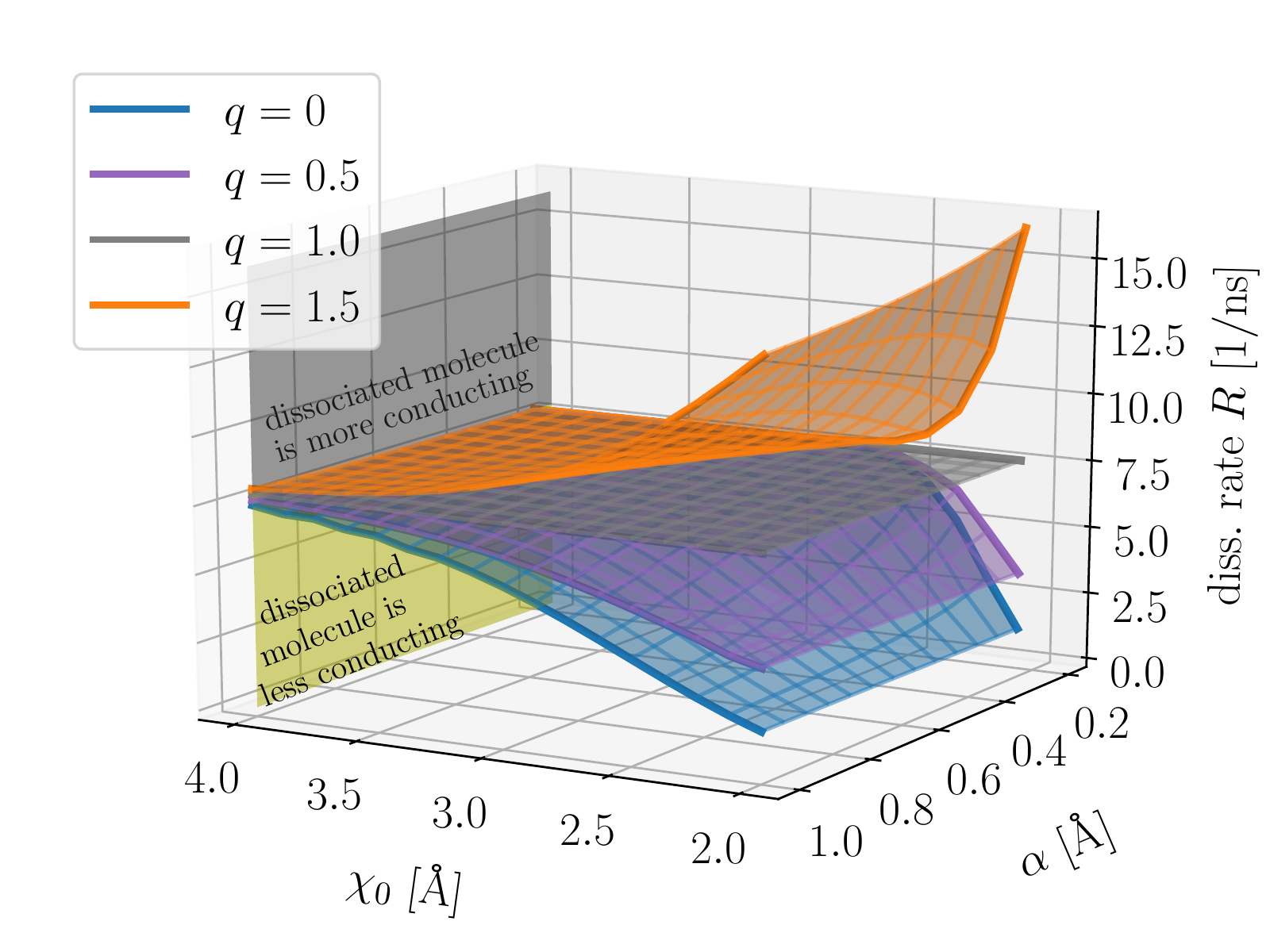}\\
			\captionsetup{font=small,labelfont=bf, justification=centerlast, format=plain}
			\caption{ \bf \scriptsize
					a: Dissociation probability as a function of time for a system symmetrically coupled to the leads with $\Gamma_{L/R}=0.1$ eV, bias $\Phi=2.0$ V, and $\alpha=0.5$ \AA. The different lines represent data for systems with different values of $\chi_0$ and $q$, whereby the latter models the change in conductance upon dissociation of the molecular junction. For reference, the case where the conductance is independent of the nuclear coordinate is given by the black line.
					b: Dissociation rate as a function of change in conductance $q$.
					c: Influence of $\chi_0$ and $\alpha$ on the stability of the molecular junction for representative values of $q$.
					For reference, the case where the conductance is independent of the nuclear coordinate is given by the gray surface.
				}
			\label{fig:tunneling}
		\end{figure}
        \begin{figure}[tb!]
			\centering
            \includegraphics[width=0.46\textwidth]{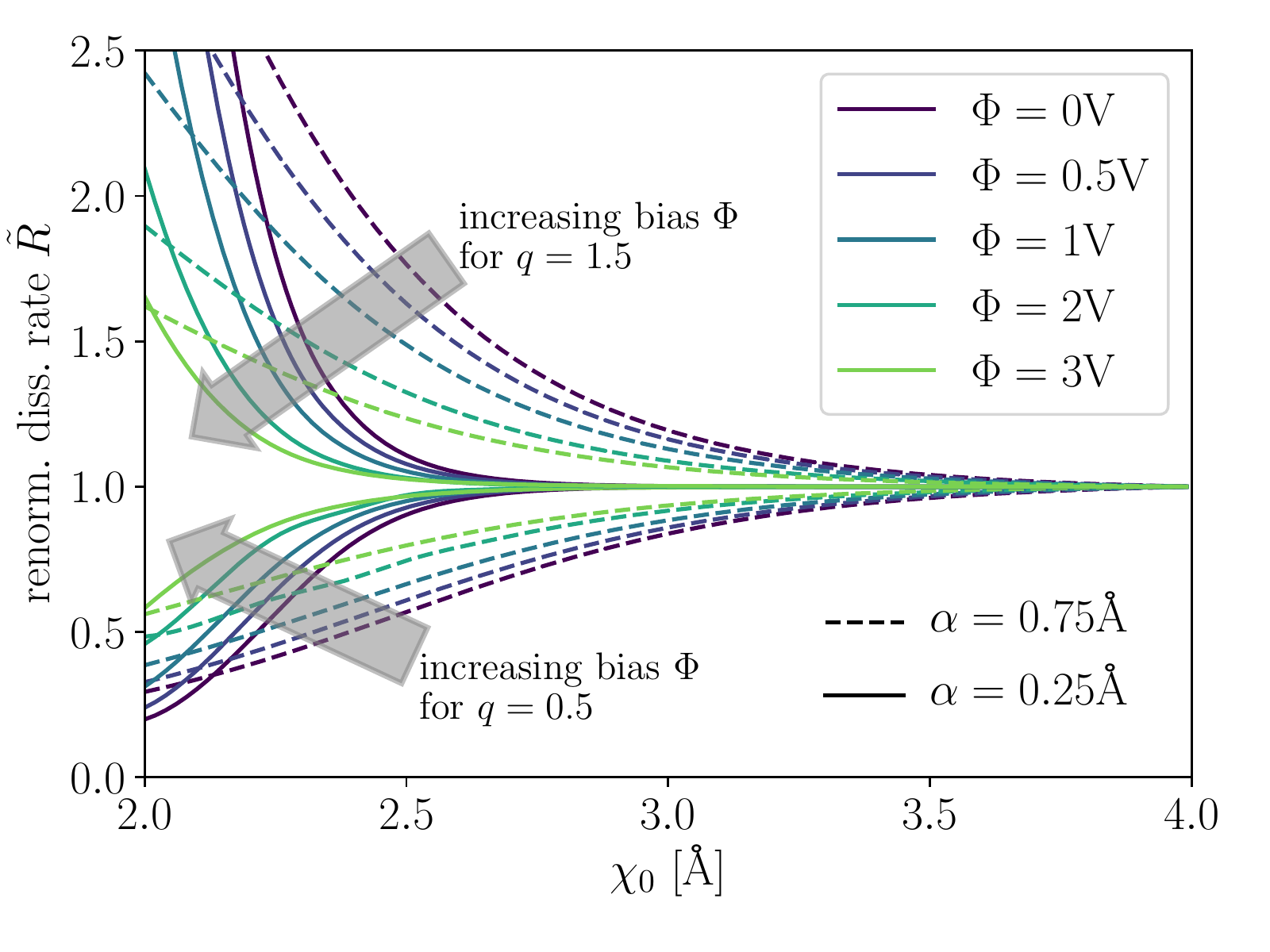}
			\captionsetup{font=small,labelfont=bf, justification=centerlast, format=plain}
			\caption{ \bf \scriptsize
					Renormalized dissociation rate $\tilde R$ as a function of $\chi_0$ for different bias voltages $\Phi$ within the tunneling regime.
					The coupling is symmetric with \mbox{$\Gamma_{L/R}=0.1$ eV}, the other relevant parameters are given in the plot.
				}
			\label{fig:tunneling_bias_dep}
		\end{figure}
		
	We begin by providing an overview of the dissociation dynamics for different representative parameters $\chi_0$ and $q$ in Fig.\ \ref{fig:tunneling}a.
	This data shows that the dissociation dynamics is highly sensitive to the different parameters modulating the change in conductance upon dissociation.
	Moreover, it establishes that in this regime, the dissociation can be described by a dissociation rate $R$. 
	In the following, we study the molecular stability as a function of the conductance change based on the dissociation rate $R$ and its dependence on $\alpha$, $q$, and $\chi_0$. Thereby, $R$ is determined as a fit parameter to the actual data. We mention that it is generally also possible to calculate the reaction rate directly within the HEOM framework.\cite{Ke_Nonequilibrium_2022}
	
	The relationship between the dissociation rate $R$ and the change in conductance upon dissociation, which is predominantly modeled by the parameter $q$, is depicted in Fig.\ \ref{fig:tunneling}b for representative values of $\alpha$ and $\chi_0$.
	Recalling that junctions whose conductivity is diminished upon dissociation are represented by values $q<1$, whereas junctions that increase their conductance upon dissociation are characterized by $q>1$, the data shows that the junction stability is enhanced for junctions that have a lower conductance in the dissociated state. This effect is more pronounced for smaller $\chi_0$ and depends only weakly on $\alpha$.
	A more detailed analysis of the dependence of the dissociation rate $R$ on $\alpha$ and $\chi_0$ is provided in Fig.\ \ref{fig:tunneling}c for representative values of $q$.
	As before, junctions with decreasing conductance upon dissociation display smaller dissociation rates, which means that they are more stable, while junctions that become more conducting upon dissociation are less stable. 
	For reference, the case where the conductance is independent of the nuclear coordinate, which is equivalent to $q=1$, is shown as a gray surface.
	The data shows that the effect of a non-constant molecule-lead coupling is most pronounced for smaller values of $\chi_0$.
	The influence of the parameter $\alpha$, which determines how rapidly the conductance changes, is to smoothen this trend.
	
	We interpret this behavior based on the underlying dissociative mechanism: Dissociation in the regime under consideration relies on the transition from the neutral to the charged state as provided by the exponentially suppressed tails of the nuclear wavefunction (see $\circled{1}$ and$\circled{2}$ in Fig.\ \ref{fig:sketch}b). The bottleneck for this process is the transition from the neutral to the charged PES, which is mediated by the molecule-lead coupling strength and correspondingly depends on the nuclear coordinate.
	For values of $q< 1$, the molecule-lead coupling is reduced, hindering the transition between neutral and charged state thus stabilizing the junction.
	This effect is most pronounced if the change in the molecule-lead coupling happens close to $x$-values that are most relevant for the transition between the charged states (for zero bias voltage, these are the $x$-values close to where $\epsilon_0(x)$ and $\epsilon_1(x)$ cross), i.e.\ when $\chi_0$ is close to this regime. 
	If $\chi_0$ is beyond this regime, the non-constant nature of the conductance can not influence the bottleneck step of the dissociative mechanism, such that the dissociation rate becomes largely insensitive to the parameters modulating the molecule-lead coupling strength. 
	For values of $q> 1$, the situation is reversed, the molecule-lead coupling strength is enhanced and so is the transition between the neutral and the charged state of the molecule, resulting in a less stable molecular junction.

	Next, we study the influence of the bias voltage. While the current in this regime does not exhibit any additional features due to the non-constant molecule-lead coupling strength (data not shown), we investigate the interplay between dissociation and applied bias voltage in Fig.\ \ref{fig:tunneling_bias_dep}.
	As the dissociation rate $R$ generally increases with applied bias voltage and spans several orders of magnitude,\cite{Erpenbeck_Current_2020} the influence of the non-constant molecule-lead coupling becomes more apparent when considering the renormalized dissociated rate $\tilde R = R / R_{\text{const}}$ rather than $R$ itself. 
	Thereby, $R_{\text{const}}$ is the dissociation rate for the same system with a constant molecule-lead coupling strength (which corresponds to $q=1$).
	Fig.\ \ref{fig:tunneling_bias_dep} therefore plots $\tilde R$ as a function of $\chi_0$ and for representative values of $q$, $\alpha$, and bias voltage $\Phi$. 
	While the overall quantitative behavior of the dissociation rate remains the same for any voltage within the low bias regime, the data reveals a shift of $\tilde R$ to smaller $\chi_0$ values with increasing bias (see arrows in Fig.\ \ref{fig:tunneling_bias_dep}).
	This shift is in line with the interpretation based on the dissociative mechanism and the transition between the neutral and the charged state. 
	As an increased bias voltage facilitates this transition for an extended range of nuclear coordinates $x$,\cite{Erpenbeck_Current_2020, Erpenbeck_dissociation_2018} $\chi_0$ must be increasingly smaller to appreciably hinder this transition.
	Notice that this also implies that the effect of a variable conductance is more important for lower bias voltages, which is the regime of interest for many experimental investigations.
	In the high bias regime, however, other effects such as the energy-dependence of the coupling to the leads have been reported to effectively stabilize molecular junctions.\cite{Gelbwaser_High_2018}
	
	We conclude by remarking that even though the results presented above have been obtained for a bandwidth of $W=100$~eV in the leads, we did not find any influence of the bandwidth on the dissociative behavior and our results are in line with results for the wide-band limit. Moreover, we did not find any characteristic relation between the dissociation dynamics and the force generated by the non-constant molecule-lead coupling strength $F_{\text{SB}}$ (data not shown).
	This is remarkable as $F_{\text{SB}}$ is proportional to the derivative of the molecule-lead coupling strength with respect to the nuclear coordinate, which changes its sign as $q$ passes through $1$, while the dissociation rate is monotonic in $q$ (see Fig.\ \ref{fig:tunneling}b).

\subsection{Strong-coupling regime -- classical limit}\label{sec:strong-coupling}

	\begin{figure}[htb!]
				\raggedright a)\\
				\raggedleft
				\vspace*{-0.6cm}
				\includegraphics[width=0.46\textwidth]{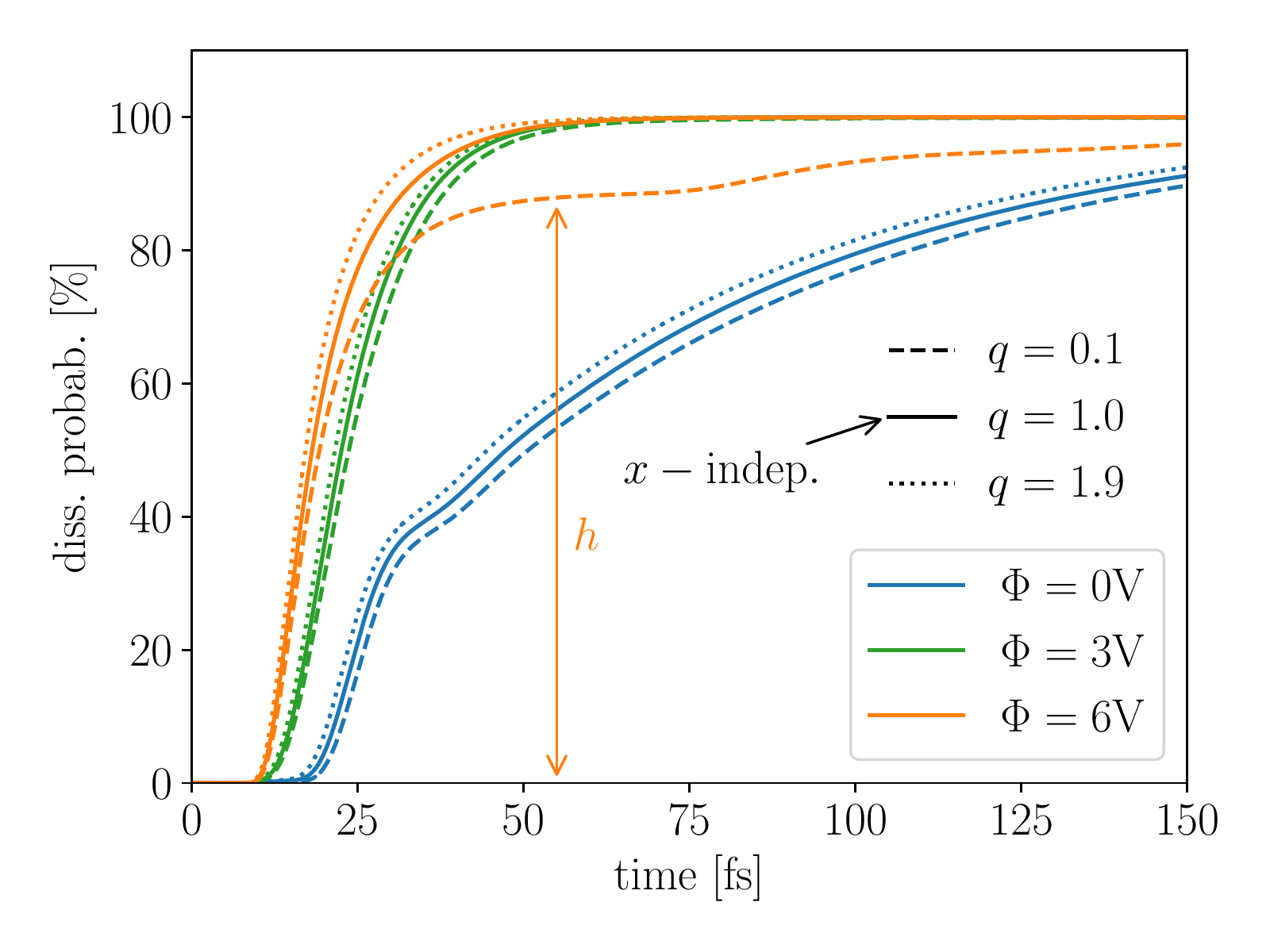}\\
				\raggedright b)\\
				\raggedleft
				\vspace*{-0.6cm}
				\includegraphics[width=0.46\textwidth]{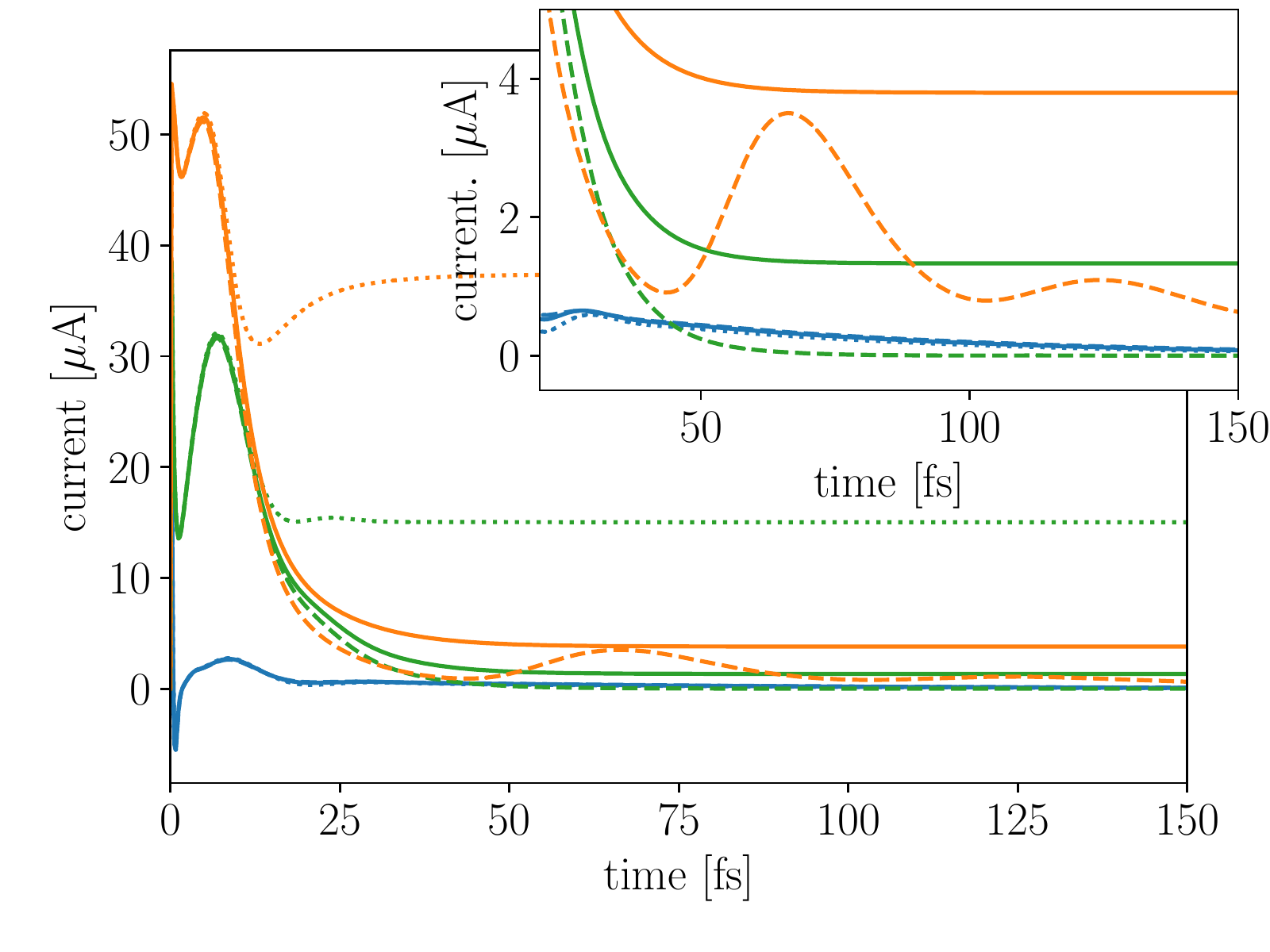}\\
				\raggedright c)\\
				\raggedleft
				\vspace*{-0.6cm}
                \includegraphics[width=0.46\textwidth]{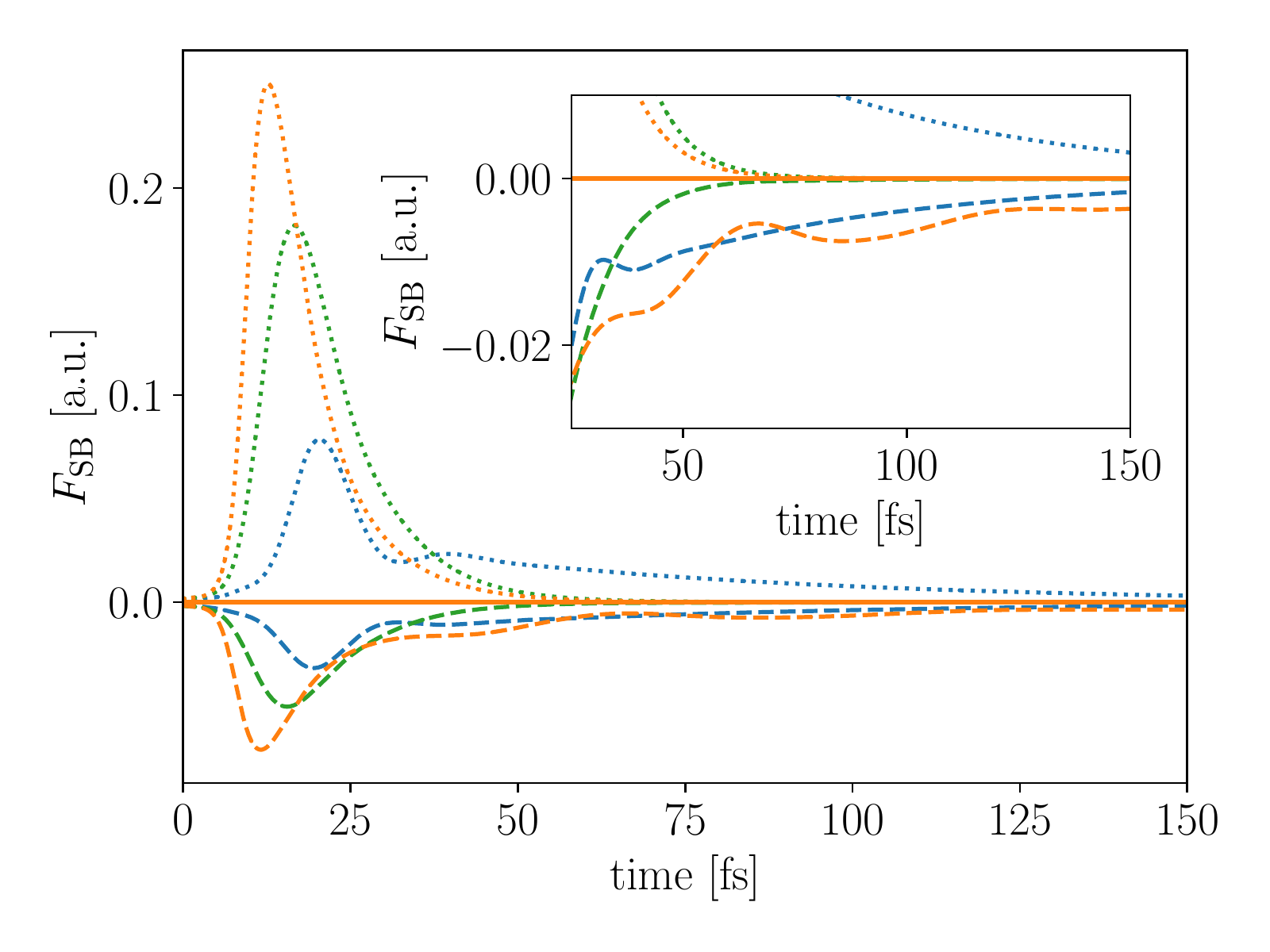}
            \captionsetup{font=small,labelfont=bf, justification=centerlast, format=plain}
			\caption{ \bf \scriptsize
                    Dynamics of a molecular junction coupled symmetrically to two leads with $\Gamma_\text{L/R}=0.5$ eV. 
					The bandwidth of the leads is $W=100$~eV.
					a: Dissociation probability as a function of time.
					The different line colors correspond to different bias voltages.
					Different line styles correspond to junctions where the dissociated molecule is a better ($q>1$) or worse ($q<1$) conductor. For reference, the limit of a nuclear coordinate independent conductance is plotted as a full line ($q=1$).
					b: Electronic current as a function of time corresponding to the data provided in subplot a. The inset provides a close-up of the long-time low current regime.
					c: Force $F_{\text{SB}}$ generated by the non-constant molecule-lead coupling as defined in Eq.\ (\ref{eq:force}) for the data provided in subplot a.
				}
			\label{fig:classical}
    \end{figure}
    \begin{figure*}[htb!]
			\begin{minipage}[c]{0.32\textwidth}
				\raggedright a)\\
				\hspace*{-1.2cm}
                \includegraphics[width=1.25\textwidth]{{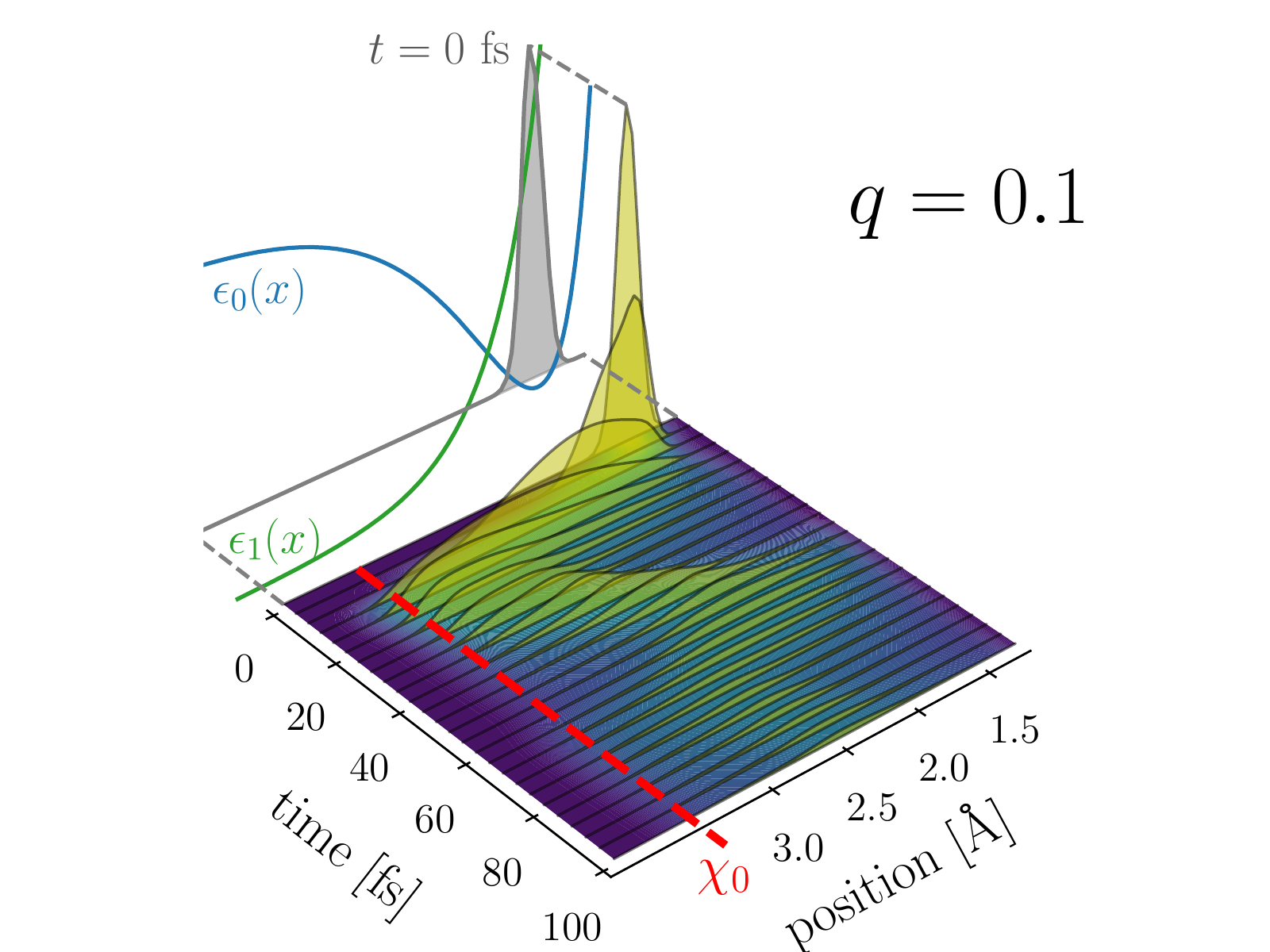}}
			\end{minipage}
			\begin{minipage}[c]{0.32\textwidth}
				\hspace*{0.3cm}\raggedright b)\\
				\hspace*{-0.7cm}
                \includegraphics[width=1.25\textwidth]{{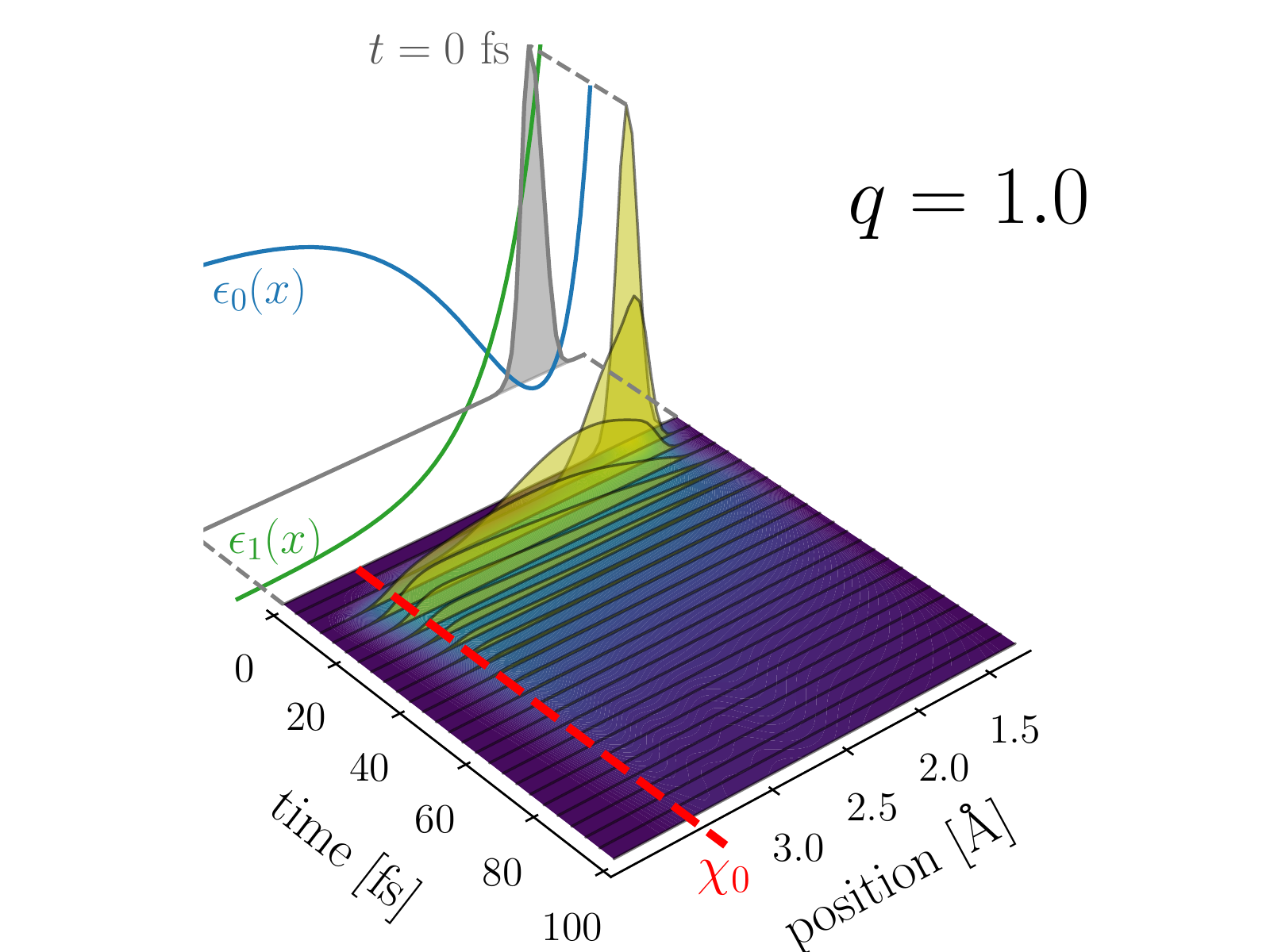}}
			\end{minipage}
			\begin{minipage}[c]{0.32\textwidth}
				\hspace*{0.6cm}\raggedright c)\\
				\hspace*{-0.4cm}
                \includegraphics[width=1.25\textwidth]{{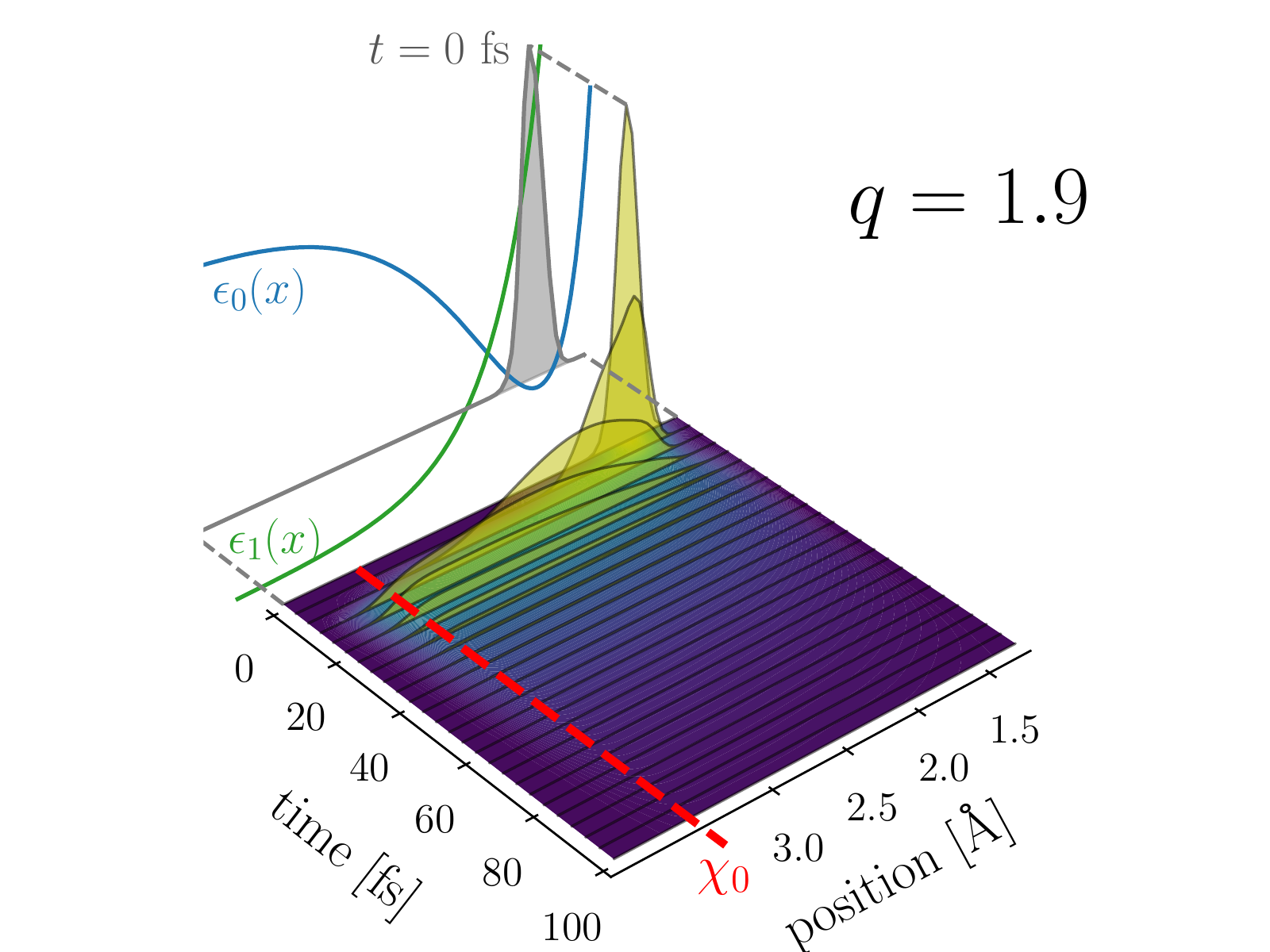}}
			\end{minipage}
			\captionsetup{font=small,labelfont=bf, justification=centerlast, format=plain}
			\caption{ \bf \scriptsize
					Dynamics of the nuclear wave-packet for molecules with different nuclear-coordinate dependent conductance. Same parameters as in Fig.\ \ref{fig:classical}.  As the effect is most pronounced for high bias voltages (cf.\ Fig.\ \ref{fig:step_height}a), we exemplify the dynamics for the very large bias voltage of $\Phi=6$~V.
				}
			\label{fig:classical_wavepaket}
    \end{figure*}

    The second part of this work focuses on the adiabatic regime of strong molecule-lead coupling.
	In this regime, the nuclear wave-packet remains localized during propagation, such that a classical interpretation of the nuclear motion becomes meaningful and the dissociation process can be interpreted in terms of forces acting on the nuclear wave-packet (Notice that this is in contrast to the weak-coupling regime, where dissociation is mainly the result of a tunneling effect, such that a classical interpretation does not provide the correct intuition.).\cite{Erpenbeck_Current_2020, Erpenbeck_dissociation_2018}
	Moreover, we recently found that under certain conditions, an increased molecule-lead coupling strength can effectively stabilize molecular junctions.\cite{Ke_Unraveling_2021} 
	
	An overview over the dissociation dynamics in the adiabatic regime is provided in Fig.\ \ref{fig:classical}a--c for representative values of $q$ and different bias voltages $\Phi$.	
    While for low to intermediate bias voltages, the overall dynamics of the system is largely independent of the conductance change $q$, the high bias voltage regime reveals the formation of distinctive steps in the dissociation probability (see Fig.~\ref{fig:classical}a).
	Interestingly, these steps only appear for $q<1$, that is, for junctions where the dissociated molecule is less conducting. 
	We find that these steps are accompanied by oscillations in the electronic current (see inset in Fig.~\ref{fig:classical}b) as well as in the force stemming from the non-constant molecule-lead coupling strength (see inset in Fig.~\ref{fig:classical}c).
	
	As a classical interpretation of the nuclear DOFs becomes meaningful in the strong coupling regime, we choose to analyze our findings based on the additional force $F_{\text{SB}}$ that is generated by the non-constant molecule-lead coupling strength (Fig.~\ref{fig:classical}c) and its influence on the dynamics of the nuclear wave-packet (Fig.~\ref{fig:classical_wavepaket}a--c).
	As $F_{\text{SB}}$ is proportional to the derivative of the molecule-lead coupling strength with respect to the nuclear coordinate (see Eq.~(\ref{eq:force})), its direction is different for situations where the conductance is increased or decreased for large $x$-values. 
	In the case that the conductance is decreased for the dissociated molecule, this additional force pushes the nuclei in the direction of their equilibrium position. Accordingly, in Fig.~\ref{fig:classical_wavepaket}a, we see that a part of the nuclear wave-packet gets reflected by this additional force and propagates back to the equilibrium position. This particular behavior results in the formation of steps in the dissociation probability, i.e.\ 
	in increasing the stability of molecular junctions where the dissociated molecule is less conducting.
	Notice that even though $F_{\text{SB}}$ is also acting on molecules that are more conducting if dissociated, this force does not lead to a stabilization in this case (see dynamics of wave-packet in Fig.~\ref{fig:classical_wavepaket}c).
	Moreover, the data shows an increase in the magnitude of $F_{\text{SB}}$ with an increase in bias voltage. The impact of  $F_{\text{SB}}$ on the system is accordingly most pronounced in the high bias regime when appreciable amounts of current flow across the molecule. 
	Due to the similarities of the expressions for the current and the force $F_{\text{SB}}$ (see Eq.\ (\ref{eq:current}) and Eq.\ (\ref{eq:force}), both depend on the first-tier auxiliary density matrices), it can be implied that the force depends on the number of electrons exchanged between the molecule and the leads.
	This makes the phenomenon an intrinsic nonequilibrium effect, which is in contrast to the behavior observed in the non-adiabatic regime, where no obvious dependency on the current was found.

    \begin{figure}[htb!]
			\raggedright a)\\
			\raggedleft
            \vspace*{-0.6cm}
			\includegraphics[width=0.46\textwidth]{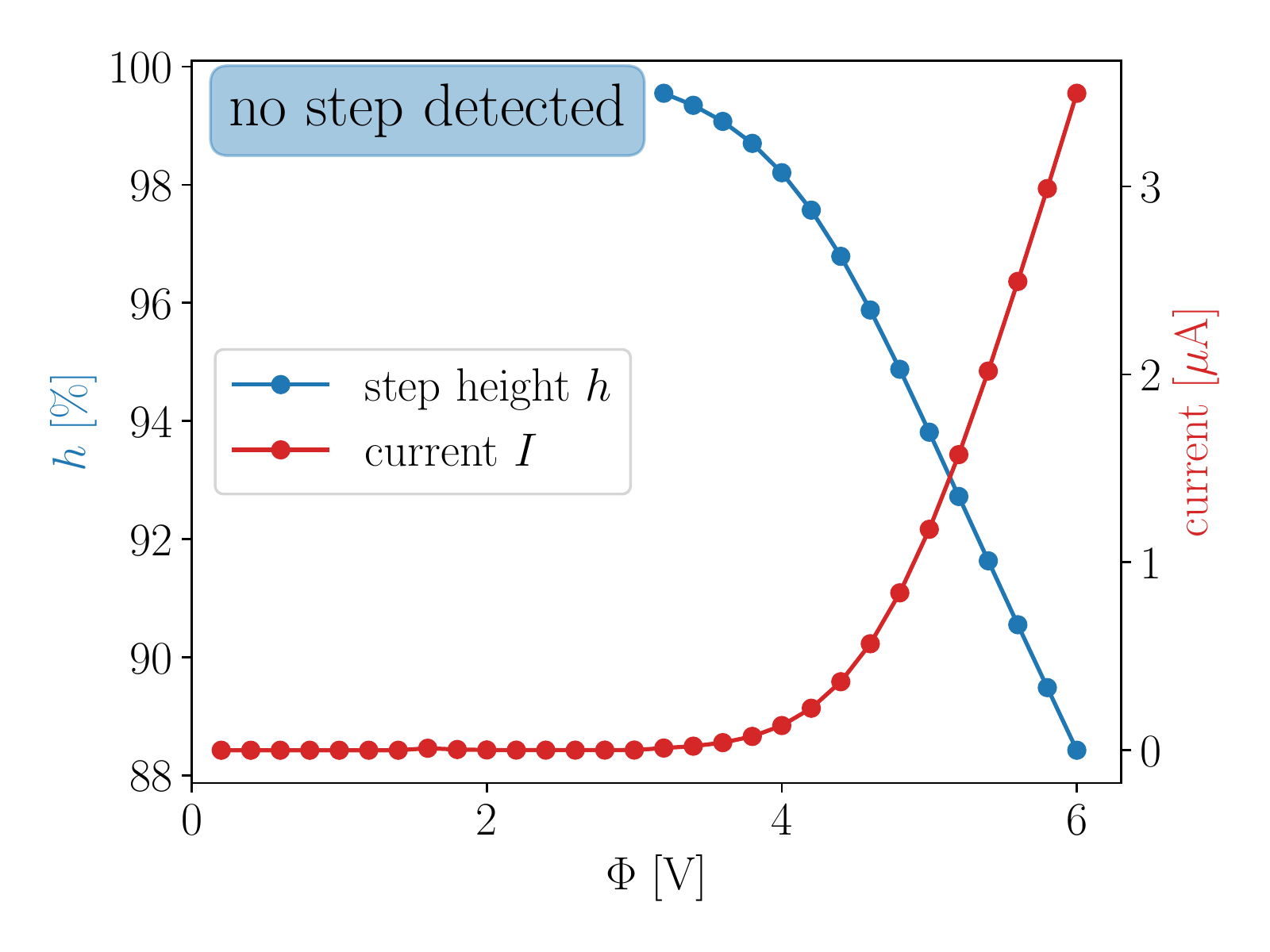}\\
			\raggedright b)\\
			\raggedleft
            \vspace*{-0.6cm}
			\includegraphics[width=0.46\textwidth]{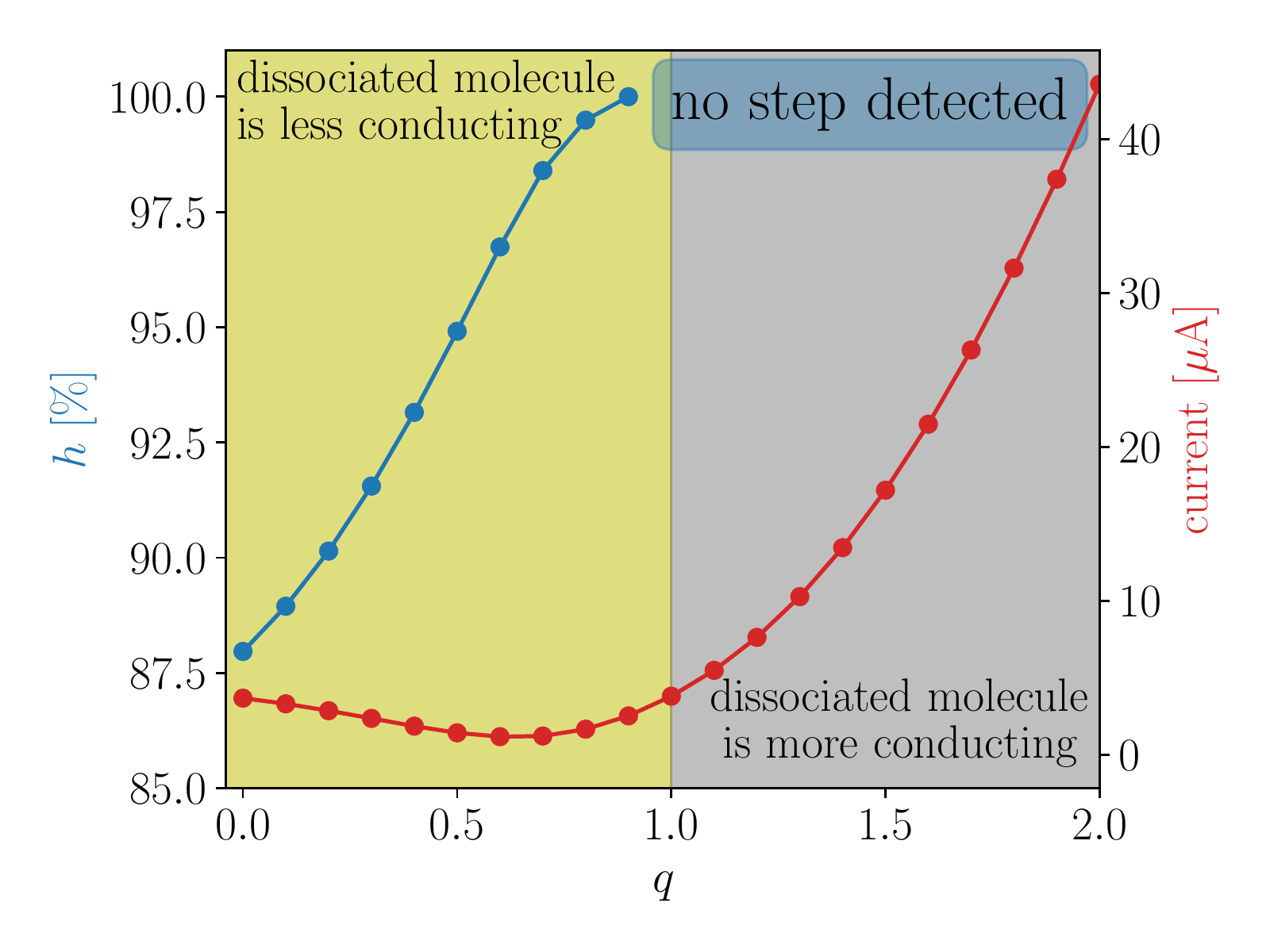}\\
			\raggedright c)\\
			\raggedleft
            \vspace*{-0.6cm}
            \includegraphics[width=0.46\textwidth]{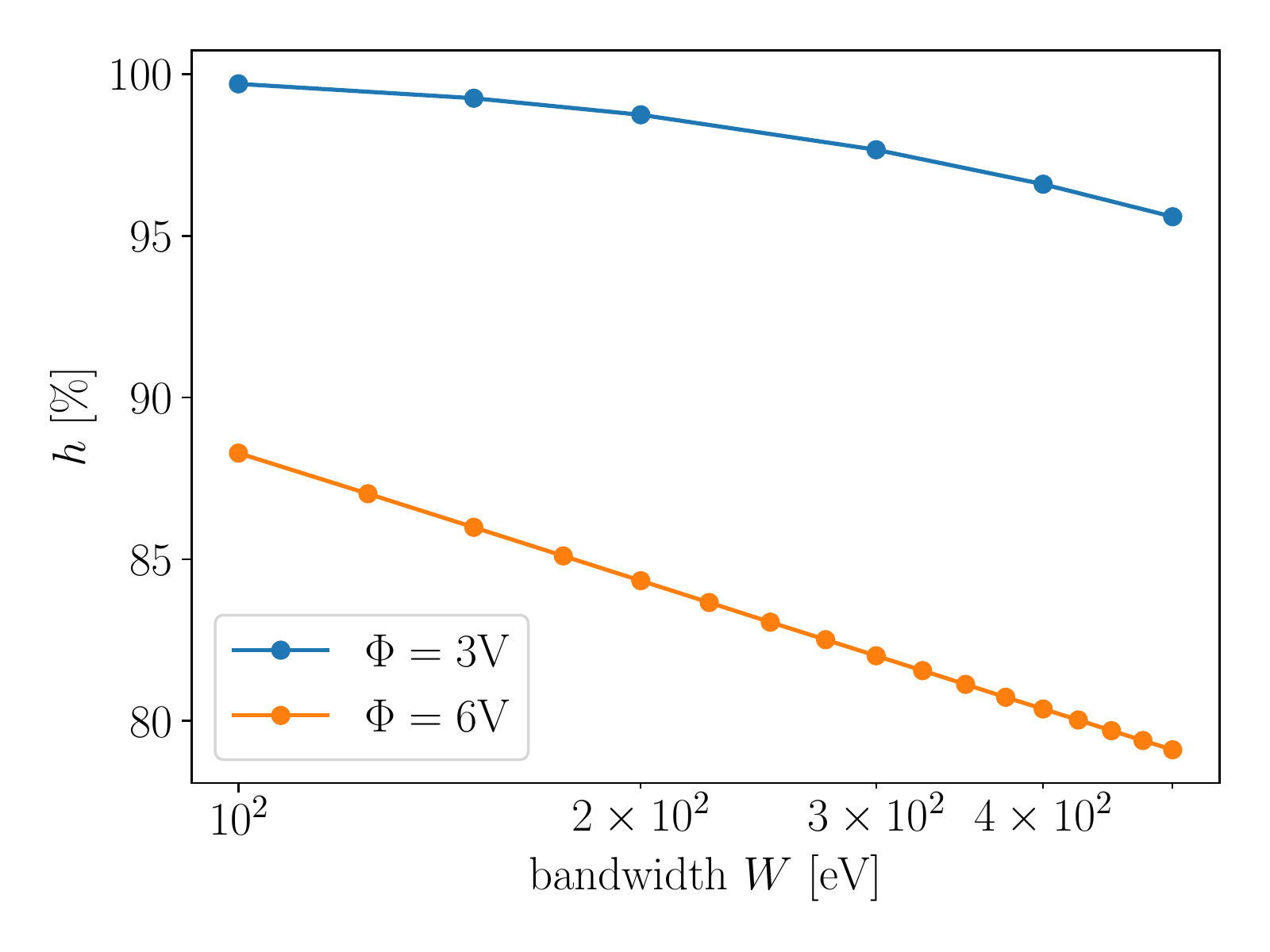}
		\captionsetup{font=small,labelfont=bf, justification=centerlast, format=plain}
		\caption{ \bf \scriptsize
                Dependence of the dissociation step height $h$ on the different parameters that determine the difference in conductance between an intact and a dissociated molecular junction.
                The molecular junction is coupled symmetrically to two leads with $\Gamma_\text{L/R}=0.5$ eV.
                As the corresponding calculations are numerically expensive, the actual data is highlighted by dots.
                a: Step height and current as a function of bias voltage for $q=0.1$ and $W=100$~eV.
                b: Step height and current as a function of change in conductance $q$ for $\Phi=6$~V and $W=100$~eV.
                c: Step height for two representative bias voltages as a function of bandwidths in the leads for $q=0.1$.
			}
		\label{fig:step_height}
    \end{figure}
    
    In the following, we systematically explore the molecular stability as a function of conductance change in the adiabatic regime. 
    To this end, we use the height of the steps $h$ in the dissociation probability (as exemplified in Fig.~\ref{fig:classical}a) as a figure of merit and study its dependence on the bias voltage $\Phi$, the change in conductance $q$, and the bandwidth in the leads $W$ in Fig.~\ref{fig:step_height}. Thereby, we determine the position of the step by finding the first minima in the derivative of the dissociation probability with respect to time, and therefrom deduce the height $h$.
    
    The influence of an applied bias voltage on the step height $h$ is investigated in Fig.~\ref{fig:step_height}a.
    We observe that there is an onset bias voltage, below which we do not find steps in the dissociation probability. Above this bias voltage, the step height depends in a monotonic way on the applied bias voltage, i.e.\ the step is smaller and the molecule therefore more stable for higher bias voltages.
    Moreover, the step height correlates with the current flowing across the molecule. 
    This is in line with our understanding that the steps are a result of the force $F_{\text{SB}}$, which is related to the number of electrons exchanged between the molecule and the leads. 
    Consequently, molecules can be more stable in the high bias regime where they are carrying larger currents, than in the lower bias regimes where they are carrying smaller currents.
    This is in line with previous works that have reported that molecular junctions can be more stable in the high bias regime in the presence of an energy-dependent molecule-lead coupling.\cite{Gelbwaser_High_2018}
    
    In Fig.~\ref{fig:step_height}b, we plot the step height $h$ as a function of difference in the conductance between the stable and the dissociated molecule as modeled by $q$.
    We observe that steps in the dissociation probability as a function of time are only found for molecules that are less conducting in the dissociated configuration, that is $q<1$. Moreover, the step heights are smaller the more $q$ deviates from $1$, that is the limiting case where the conductance is independent of the nuclear coordinate.
    This implies that junctions that are bad conductors if the molecule is dissociated, are generally more stable.
    As before, we find that the trend in the step height and the associated trend in the current agree with our interpretation that the effect is based on the force $F_{\text{SB}}$, whereby both the current and the force $F_{\text{SB}}$ can be expressed in terms of first-tier auxiliary density operators (compare Eq.(\ref{eq:current}) and Eq.\ \ref{eq:force}).
    
    We investigate the influence of the bandwidth $W$ in the leads on the stability of the junction as given by the step height $h$ in Fig.~\ref{fig:step_height}c.
    We observe that the  step height $h$ decreases with an increase of $W$. For $\Phi=6$~V, we observe an exponential decay of the step height with $W$, whereas the decay is roughly linear in $W$ for $\Phi=3$~V.
    This behavior suggests that junctions are more stable if the molecule is attached to leads with a larger bandwidth.
    This is a remarkable observation, as it suggests that the junction stability is also influenced by properties of the electrodes, i.e.\ the stability of a molecular junction does not only depend on the properties of the molecule. The influence of the leads on the junction stability was previously only discussed based on mechanisms that effectively cool the molecular junction.\cite{Gelbwaser_High_2018}

\section{Conclusion}\label{sec:conclusion}

    The stability of molecular junctions is fundamental for the field of molecular electronics.
    An inherent feature of molecular junctions at the limit of mechanical stability is a change in their conductance.
    In this work, we investigated the influence of current-induced changes in the conductance on the stability of molecular junctions.
    To this end, we applied a numerically exact framework based on the HEOM method to generic models for molecular junctions with nuclear coordinate dependent molecule-lead coupling strengths.
    The central finding of this work is that molecular junctions that exhibit a decrease in conductance upon dissociation are more stable than junctions that are more conducting in their dissociated state.
    We have identified different stabilization effects 
    that explain this finding, whereby
    the underlying mechanism depends on the transport regime and is very different for the adiabatic and the nonadiabatic limit. 
    In the adiabatic regime where a classical interpretation based on forces acting on the nuclei becomes meaningful, we find that the stabilization effect is related to the current flowing across the junctions and is thus most pronounced in the high bias regime.
    Further, we find unique signatures in the current associated to this stabilization effect, that depend on the bandwidth in the leads.
    These observations are essential for the design of molecular electronics as they provide guidelines for constructing stable molecular junctions which are not only based on properties of the molecule, but also take into account aspects of the leads.

\section*{Acknowledgements}
	This work was supported by a research grant (TH 867/8-1) of the German Research Foundation (DFG).
	A.E.\ was funded by the German Research Foundation -- 453644843.
	U.P.\ acknowledges support from the Israel-US bi-national science foundation, grant 2020327.
	This research used resources of the National Energy Research Scientific Computing Center, a DOE Office of Science User Facility supported by the Office of Science of the U.S. Department of Energy
    under Contract No. DE-AC02-05CH11231 using NERSC award BES-ERCAP0021805.
    The authors acknowledge support by the state of Baden-Württemberg through bwHPC and
    the German Research Foundation (DFG) through Grant No. INST 40/575-1 FUGG (JUSTUS 2 cluster).

\bibliography{Bib}
\end{document}